\documentclass[sigconf,manuscript]{acmart}
\makeatletter
\renewcommand\footnotetextcopyrightpermission[1]{} 
\AtBeginDocument{%
  \providecommand\BibTeX{{%
    Bib\TeX}}}

\acmISBN{978-1-4503-XXXX-X/18/06}

\usepackage{array}
\usepackage{graphicx}
\usepackage{booktabs}
\usepackage{multirow}
\usepackage{threeparttable}
\usepackage[utf8]{inputenc} 
\usepackage[T1]{fontenc}      
\DeclareUnicodeCharacter{FFFD}{\textquestiondown}

\begin{document}
\title{When Trust Collides: Decoding Human-LLM Cooperation Dynamics through the Prisoner’s Dilemma}

\author{Guanxuan Jiang\textsuperscript{1}, Shirao Yang\textsuperscript{1}, Yuyang Wang\textsuperscript{1,*}, Pan Hui\textsuperscript{1,2}}
\affiliation{\textsuperscript{1}%
  \institution{The Hong Kong University of Science and Technology (Guangzhou)}
  \city{Guangzhou}
  \country{China}}
\affiliation{\textsuperscript{2}%
  \institution{The Hong Kong University of Science and Technology}
  \city{Hong Kong}
  \country{China}}
\email{{gjiang240, syang610, yuyangwang}@connect.hkust-gz.edu.cn, panhui@ust.hk}
\footnote{* Corresponding Author}

\begin{abstract}

As large language models (LLMs) become increasingly capable of autonomous decision-making, they introduce new challenges and opportunities for human-AI cooperation in mixed-motive contexts. While prior research has primarily examined AI in assistive or cooperative roles, little is known about how humans interact with AI agents perceived as independent and strategic actors. This study investigates human cooperative attitudes and behaviors toward LLM agents by engaging 30 participants (15 males, 15 females) in repeated Prisoner’s Dilemma games with agents differing in declared identity: purported human, rule-based AI, and LLM agent. Behavioral metrics—including cooperation rate, decision latency, unsolicited cooperative acts and trust restoration tolerance—were analyzed to assess the influence of agent identity and participant gender. Results revealed significant effects of declared agent identity on most cooperation-related behaviors, along with notable gender differences in decision latency. Furthermore, qualitative responses suggest that these behavioral differences were shaped by participants’ interpretations and expectations of the agents. These findings contribute to our understanding of human adaptation in competitive cooperation with autonomous agents and underscore the importance of agent framing in shaping effective and ethical human-AI interaction.
\end{abstract}

\ccsdesc[500]{Human-centered computing~Human computer interaction (HCI)}
\ccsdesc[500]{Human-centered computing~Collaborative and social computing}
\ccsdesc[500]{Computing methodologies~Artificial intelligence}

\keywords{Large language model, Game theory, Human-AI interaction, Human behavior, Social cooperation}

\maketitle
 
\section{Introduction}
Over the past five years, the field of artificial intelligence (AI) has experienced significant growth, largely driven by the rapid advancement and deployment of large language models (LLMs) \cite{wu2023brief}. At the same time, with the influx of autonomous decision-making AI agents, interactions between humans and AI agents are becoming increasingly complex. Meanwhile, existing research on human-AI interaction has primarily focused on AI as assistive tools that augment human capabilities under human supervision \cite{xu2021human, amershi2019guidelines, 10.1145/3579612}. From a longer-term perspective, recent advances in LLMs have enabled AI agents to execute complex tasks autonomously—from strategic planning to conflict resolution \cite{luo2024large}. Given that more and more research underscores the effectiveness of LLMs in simulating human cognition and cooperative behavior \cite{ke2024exploring, abbasiantaeb2024let}, these models hold the potential to integrate into human society as quasi-autonomous partners, capable of making independent decisions during interactions with humans and thereby enhancing performance across various domains \cite{liu2023summary, gong2025effectivefixedtimecontrolconstrained}.

Despite these promising developments, human attitudes toward cooperation with LLMs remain ambiguous due to various factors. Human attitudes and behaviors toward AI agents are strongly influenced by how AI is perceived, particularly when AI transitions from a tool to an independent decision-making entity \cite{fahnenstich2024trusting}. This transition fundamentally changes human-AI interaction by introducing negotiation and trust-building processes traditionally reserved for human-to-human collaboration \cite{duan2025understanding}. Autonomous AI entities challenge norms in cooperative tasks by requiring humans to evaluate the AI’s decision-making for accuracy and alignment with human goals and values. However, when humans are faced with LLMs as autonomous partners, concerns about the “black-box” nature of these systems—that is, the opacity and difficulty in interpreting their decision-making processes \cite{nguyen2023black, Chkirbene2024Large}—further exacerbate skepticism in critical decision-making scenarios. 

Humans' recognition of AI agents' identities, along with variations in their trust attitudes towards AI, plays a crucial role in shaping cooperative decision-making. On the one hand, how AI agents present their identity—whether explicitly labeled as non-human or intentionally designed to mimic human-like traits—can significantly influence user attitudes, trust levels, and cooperative behaviors during interactions \cite{geiselmann2023interacting}. For example, humans tend to demonstrate greater trust and willingness to cooperate with transparent systems like rule-based AI, whereas opaque systems such as LLMs often elicit more cautious or adversarial behaviors, particularly in high-stakes collaborative scenarios \cite{lamparth2024human}. Besides, due to the perception and interaction styles, gender-specific variations in trust in AI agents have emerged as a critical factor in shaping the cooperation results in human engagement with AI agents \cite{jermutus2022influences,varona2022discrimination}.

To further explore how autonomous decision-making AI agents reshape interaction dynamics, it is imperative to figure out several key factors that impact humans adapting their cooperative strategies based on the AI's perceived agency and trust intention. Based on the research review, our study focuses on two perspectives: the declared identity of the LLM agent and the human gender. Additionally, considering the practical implications for improving cooperative interactions, we emphasize considering and deconstructing the contextual factors that affect trust dynamics, communication, and outcomes during this cooperative decision-making process. To address these issues, we employed the classic ``Prisoner's Dilemma'' as an experimental framework and proposed the following research questions:

\begin{itemize}
\item \textbf{RQ1:} How does the declared identity of the LLM agents influence human's cooperative attitude and behavior in prisoner's dilemmas?
\item \textbf{RQ2:} How does gender influence human's cooperative attitude and behavior towards LLM agents in prisoner's dilemmas?
\item \textbf{RQ3:} What other contextual factors mediate the cooperative attitude and behavior between humans and LLM agents?
\end{itemize} 

This study makes four key contributions to the field of Human-AI Interaction. First, it demonstrates that the declared identity of AI agents—whether human, rule-based, or LLM—significantly affects user cooperation rates, decision latency, and trust tolerance in repeated interactions. Second, it identifies gender-related differences in decision latency, suggesting that participant gender plays a role in shaping temporal aspects of decision-making with AI partners. Third, it investigates the interaction effects between agent identity and participant gender to uncover the factors driving these behavioral disparities. Finally, it provides empirical evidence of how humans adapt their strategies in mixed-motive interactions with autonomous AI agents, advancing our understanding of interaction dynamics beyond purely cooperative paradigms.

\section{Related Work}

\subsection{Human-AI Cooperation in the Prisoner's Dilemma}

Game theory provides a mathematical framework for analyzing decision-making scenarios where multiple agents influence outcomes, with the Prisoner's Dilemma being a key model for studying cooperation \cite{branzei2008models, rajeswaran2020game}. Recently, game theory has been applied in computational sociology to study cooperation among agents, with a growing focus on human-AI interactions within the Prisoner's Dilemma \cite{xu2020game, jain2020recognition, yang2020overview, gradin2016neural}. \citet{oskamp1971effects} and \citet{haesevoets2018behavioural} have both highlighted that the mixed-motive scenarios presented in the prisoner’s dilemma serve as an effective measurement for examining social interaction issues within complex motivational environments.

\citet{fontana2024nicerhumanslargelanguage} employs the Prisoner's Dilemma to examine the cooperative tendencies and behavioral patterns of various LLMs over repeated interactions. This work introduces a systematic framework for analyzing LLM decision-making in social contexts, contributing to research on AI auditing and alignment. \citet{ng2023communicative} explores human cooperative behavior with different types of partners (human, cooperative AI, non-cooperative AI) through the framework of the Prisoner's Dilemma. This work highlights the value of the Prisoner's Dilemma in studying human-AI collaboration dynamics.

In the study of human behavior within the Prisoner's Dilemma, numerous variables have been proposed as indicators to assess interaction, collectively offering valuable insights into how individuals navigate strategic dilemmas. These indicators, ranging from cooperation rates to reaction times, reveal the interplay between rational deliberation and social strategies in shaping human decision-making. For example, \citet{wang2020long} identified decision latency in iterative Prisoner's Dilemma games as a key metric of deliberative processes, reflecting the extent to which participants carefully consider their opponent's strategy, which ultimately influences the overall rate of cooperation. Additionally, \citet{wang2020long} introduced unconditional cooperation—defined as unsolicited cooperative acts—as another behavioral measure. They found that individuals are inclined to reciprocate proactive cooperation and often make attempts to establish mutual cooperation before unilateral defection occurs, highlighting the importance of proactive strategies in fostering cooperation. Furthermore, forgiveness strategies offer an effective means of restoring cooperative relationships. Specifically, \citet{Martinez-Vaquero2015Apology} demonstrated that forgiving an opponent's betrayal enables participants to achieve higher payoffs over the course of long-term interactions. Moreover, the effort expended by opponents to seek forgiveness after a betrayal serves as a valuable metric for assessing behavioral decision-making within the context of the Prisoner's Dilemma. Together, these studies illustrate that decision-making in the Prisoner's Dilemma is not only influenced by rational calculation but also shaped by emotional and social dynamics inherent in human interactions.

As AI agents increasingly function as independent decision-makers, their behavior closely aligns with the strategic dynamics of the Prisoner's Dilemma. This makes the model particularly suitable for exploring cooperation, trust, and forgiveness in human-AI relationships. 
\label{sec:Human-AI Cooperation in the Prisoner’s Dilemma}
\subsection{Effects of AI Identity on Human Cooperation Dynamics}
\subsubsection{AI Identity Transparency Enhances Human Trust and Cooperation}


As AI technology proliferates across industries, human-AI interactions are becoming increasingly complex with the influx of autonomous decision-making agents. Large Language Models (LLMs), representing a significant advancement in AI, enable LLM-driven agents to comprehend and generate human-like text, facilitating seamless interactions across diverse contexts \cite{chang2024survey}. Characterized by rapid response times and high adaptability, these models foster increased user trust and willingness to engage in collaborative tasks \cite{kirk2024benefits}. By reducing interaction costs, they thereby expand avenues for human collaboration with intelligent agents \cite{kim2024understanding,liao2024effective}.

While LLM-driven agents demonstrate remarkable capabilities in human-like communication, a critical aspect of their deployment is how their identity as AI agents is disclosed or represented to users. The way AI agents present their identity—whether explicitly labeled as non-human or intentionally designed to mimic human-like traits—can significantly influence user attitudes, trust levels, and cooperative behaviors during interactions \cite{geiselmann2023interacting}. Previous studies suggest that when AI agents transparently disclose their identity, users may exhibit higher levels of trust and willingness to collaborate, as transparency fosters confidence in the agent's intentions \cite{Saffarizadeh2024"My}. Additionally, \citet{von2021transparency} proved that transparency is a necessary condition for trust and eventually for judging AI to be trustworthy by philosophical analysis of trust. Conversely, when AI agents without disclosing their non-human identity, users may initially demonstrate high engagement \cite{10.1016/j.chb.2024.108448}. However, such interactions risk undermining trust if users later perceive the agent's behavior as deceptive or misleading, highlighting the potential ethical and practical challenges in designing AI systems with varying levels of identity transparency \cite{li2024impact}.

Despite these insights, current research primarily focuses on AI agents in general, without systematically addressing how different types of AI agents—such as purported real human, rule-based algorithms, and LLM-driven agents—interact with users under varying identity disclosure conditions. This lack of differentiation overlooks the possibility that user perceptions, attitudes, and trust dynamics may vary significantly across these distinct types of AI agents.

\subsubsection{Human Cooperative Attitude differs between rule-based systems and LLMs}


Existing research highlights that human attitude and cooperation vary significantly depending on AI characteristics such as interpretability and transparency \cite{wang2023impact}. Humans may demonstrate greater trust and willingness to cooperate with transparent systems like rule-based AI. In contrast, opaque systems such as LLMs often elicit more cautious or adversarial behaviors, especially in high-stakes collaborative scenarios \cite{lamparth2024human}. For instance, rule-based systems are often preferred for their predictability and interpretability, fostering trust in contexts requiring accountability \cite{rane2023explainable}. Conversely, LLMs, despite their human-like communication abilities, are perceived as ``black-box'' systems due to their lack of transparency \cite{Chkirbene2024Large}, which can lead to skepticism in critical decision-making scenarios. 

These differences in system design influence trust and cooperative behavior: rule-based systems can facilitate stable and predictable collaboration, whereas LLMs can foster emotional connection while introducing uncertainty \cite{lu2024llms}. For instance, \citet{panagoulias2023rule} emphasizes the lack of trust in LLMs in complex and high-stakes decision-making scenarios, such as in healthcare. He suggests that combining rule-based systems with LLMs can enhance their credibility. This also suggests that while LLMs excel in generalization, they may disrupt cooperation in applications requiring high transparency or accountability. These differences influence user attitudes and cooperation behavior across different contexts, reflecting the importance of AI type in shaping human behavior. 

In summary, existing research primarily examines the applicability of different AI types across various contexts, often focusing on trust and usability issues while treating AI as tools or passive assistants. This perspective fails to address how AI agents that make autonomous decisions shape cooperation dynamics. Such agents introduce nuanced patterns that require humans to adapt their cooperative strategies based on the AI's perceived agency, which are critical for optimizing human-AI collaboration, as they directly impact decision outcomes and negotiation processes.

\subsection{Human Gender Matters in Cooperation Attitudes with AI Agents}

While existing research has established the effects of AI identity representation and system types on human cooperation\cite{ferrario2020ai,pratchett1999new,kelly2023factors,xu2025enhancing}, individual differences like gender remain underexplored in shaping these dynamics. Especially, gender-specific variations in trust, perception, and interaction styles have emerged as critical factors in shaping how humans engage with AI agents, highlighting the need for a deeper understanding of these dynamics \cite{jermutus2022influences,varona2022discrimination}.

\citet{ahn2022effect} found that female participants are generally more inclined than male participants to trust and accept AI recommendations in consumer contexts. This suggests that females prefer to view AI as an assistant that helps them identify the necessary products. At the same time, \citet{ofosu2023gender} found that gender significantly influences STEM students’ acceptance and use of AI-based tools in collaborative tasks. Due to their comparatively lower interest in technology, female students are more reluctant to use AI tools, limiting their opportunities to leverage AI assistance to enhance their academic performance. \citet{renz2024me} reported that females are less willing to use Strong AI than males, a disparity that is influenced by their emotional responses toward the technology.

These conclusions collectively demonstrate that which gender the participant belongs to may play a significant role in human-AI interactions and collaborative processes because of the emotional responses and influencing human behavior and collaborative performance.

In summary, prior studies highlight the significance of gender in shaping attitudes toward and acceptance of AI systems, which concentrated on individual preferences or decision-making contexts, leaving the dynamic nature of human-AI collaboration largely unexamined.

\section{Method}
To address the research questions outlined above, we experimented to compare the cooperation attitude and behavior of humans towards LLM agents with different declared identities. Based on the specific LLM agents' declared identities and related settings, we conducted the formal experiment enrolling 30 participants after a pilot study for exploration. Within the framework of our experimental design, two independent variables were carefully defined: the \textit{declared identity} of the AI agent, which included purported human, purported rule-based AI agents, and LLM-based AI agents, and the \textit{gender} of the participants, categorized as male or female.

\subsection{Pre-experimental Setting}

\subsubsection{Declared Identity Setting}
We devised the following experimental scenario to investigate human cooperative behavior patterns when interacting with LLM agents functioning as potential independent decision-makers in human society. The experiment consisted of three sessions. Before each session, participants were informed about the identities of their opponents, which were, respectively, a purported human, a purported rule-based AI agent, an LLM agent. However, the actual situations are in the Table ~\ref{table:experimental_design}. 

\begin{center}
\begin{table}[htbp]
\small
\caption{Experimental Opponents' Settings}
\label{table:experimental_design}
\begin{tabular}{p{0.3cm} p{3cm} p{2.8cm} p{1cm}}
    \toprule
    No. & Purported Role & Actual Backend & Rounds \\ 
    \midrule
    1 & Rule-Based AI Agent & LLM Agent & 50 \\ 
    2 & LLM Agent & LLM Agent & 50 \\ 
    3 & Real Human & LLM Agent & 50 \\ 
    \bottomrule
\end{tabular}
\end{table}
\end{center}
\vspace{-2em}
  
For situation 1, the core description of the rule-based AI agent highlighted that is based on human-coded rules and algorithms to guide their behavior in gameplay. Expert programmers carefully design these algorithms to ensure the agent's decision-making process is interpretable, predictable, and grounded in logical reasoning \cite{russell2016artificial}. However, we cannot disclose the underlying logic and code structure to participants. For situation 2, the agents driven by LLMs were emphasized that these models were neither pre-trained for the experiment nor subjected to human manipulation. Additionally, we informed participants that their LLM agent opponents are based on the ChatGPT API\footnote{\url{https://www.openai.com}}, specifically using the gpt-4-turbo-20240409 version. This emphasis was intended to assure participants that no biases or prior knowledge bases were imposed to enhance the agents' understanding of the game. Besides, \citet{bender2020climbing} highlights that fine-tuning is often specific to particular experimental contexts and can easily result in model overfitting. Given that our research emphasizes human behavior, utilizing an unfine-tuned model is more suitable to ensure the reliability and generalizability of our findings. Furthermore, it aims to replicate real-world interactive environments, where pre-training LLM agents for every possible scenario is impractical \cite{schick2020exploiting}. 

For situation 3, as a baseline setting for comparing humans' attitude towards LLM agents, participants were explicitly informed that their opponent represented a real human player in this experimental condition. No demographic particulars (including gender, age, or social identity) were disclosed to preserve anonymity and prevent potential interpersonal bias. The human experimenter assigned to simulate this role maintained complete anonymity throughout the study - no physical, verbal, or digital interaction occurred between participants and the role-playing researcher at any experimental phase (pre-test, during, or post-test). 

\subsubsection{LLM Agents Settings}

ChatGPT API was utilized in all experimental trials as the backend, specifically leveraging the capabilities of GPT-4 Turbo\footnote{\url{https://www.openai.com}} (version gpt-4-turbo-20240409, updated on April 9, 2024). \textcolor{black}{As previously discussed, the rationale for selecting the ChatGPT API has been detailed in the preceding sections. All agents were given the game rules and scoring guidelines as default prompts. The primary objective for the agents, paralleling that of the human participants, was to maximize self-scores. This approach is grounded in value alignment theory \cite{russell2019human}, which emphasizes ensuring that agents share the same goals as participants within our experimental context. Additionally, our experiment aims to measure cooperative and non-cooperative behaviors within a competitive environment. Another justification for this setup is to prevent the introduction of extraneous variables that might confound the experimental findings.}

After each mini-round, researchers provided participants' choices to the agents. \textcolor{black}{This design enhances interactivity by offering immediate feedback, allowing agents to analyze previous data and make informed decisions. Such real-time updates enable agents to dynamically adjust their strategies based on participants' actions, fostering a more responsive and adaptive interaction. This approach aligns with human-AI interaction principles, which stress the importance of mutual behavioral understanding \cite{amershi2019guidelines}.} This feedback process is a pivotal component of the experimental design, enabling agents to analyze their previous decisions retrospectively.

\subsubsection{Game Rules}
All participants should complete the prisoner's dilemma game three times in in-person sessions based on a random sequence, with LLM agents with different purported roles stated in Table~\ref{table:experimental_design}; each consisted of 50 successive rounds, totaling 150 rounds. Based on the original rules of the prisoner's dilemma, we adapted the model to incorporate a rewarding mechanism, as illustrated in Table~\ref{table:prisoners_dilemma}. Participants were expected to maximize their scores during the game.

\begin{table}[htbp]
    \small
        \caption{Prisoner's Dilemma Payoff Matrix}
        \label{table:prisoners_dilemma}
       \begin{tabular}{p{1.5cm}p{1.3cm}p{1.5cm}p{1.3cm}}
        \toprule
        & & \multicolumn{2}{c}{Player B} \\ \cmidrule(r){3-4}
        & & Cooperate & Defect \\ \midrule
        \multirow{2}{*}{Player A} & Cooperate & (3, 3) & (0, 5) \\ \cmidrule(lr){2-4}
        & Defect & (5, 0) & (1, 1) \\ 
        \bottomrule
    \end{tabular}
        \vspace{1pt}
        \caption*{\footnotesize 
         When both players cooperate \((\text{Cooperate, Cooperate})\), the payoff is \((3, 3)\), meaning both players receive three points.
         
         When both players defect \((\text{Defect, Defect})\), the payoff is \((1, 1)\), meaning both prisoners receive one point. 
         
         If player A cooperates and player B defects \((\text{Cooperate, Defect})\), the payoff is \((0, 5)\), where player A receives nothing and player B receives five points. 
         
         If player A defects and player B cooperates \((\text{Defect, Cooperate})\), the payoff is \((5, 0)\), where player A receives five points and player B receives nothing. 
        }
        \end{table}
\vspace{0em}

This adaptation retains the same payoff structure for all four times of the game. The total score \(S\) of 150 rounds for each participant was calculated and multiplied by a fixed coefficient \(k\) to arrive at the compensation cost \(C\). Specifically, for each participant \(i\), the compensation cost \(C_i\) was calculated as follows:
\[
C_i = k \times S_i
\]
This method ensured that all participants could receive real cash compensation based on their actions, thereby motivating them to engage earnestly with the game. Furthermore, this approach more accurately simulates the incentives and benefits associated with real-world scenarios, thereby enhancing the ecological validity of the experiment. 

To ensure that participants fully comprehended the scoring rules, detailed instructions were provided for them to read before the commencement of the experiment. To verify their understanding, the trial began with two hypothetical scenarios, and participants were asked to calculate the points awarded in each situation. The experiment officially commenced after all participants correctly answered these preliminary questions. 

\subsubsection{Pilot Study}
To ensure the accessibility and robustness of the experimental design, a pilot study was conducted with 8 participants (4 males and 4 females) prior to the formal experiment. Preliminary data analysis from the pilot revealed a potential influence of gender and the AI agents' purported characteristics on cooperative behavior, which ensured the distinct feature of the following experiment findings to some extent. Based on these findings, the formal experiment was refined, and a new cohort of participants was recruited. To enhance accuracy, participants were provided with blank sheets of paper to record their choices during the games, ensuring reliable recall during subsequent questionnaires and interviews. These adjustments were intended to improve participant engagement while bolstering the validity and reliability of the experimental outcomes.

\subsection{Formal Experiment}
\subsubsection{Participants}


Participants with experience using LLM-supported AI agents (e.g., code assistants and chat boxes) were recruited for the experiment. Eventually, we have thirty (15 males, 15 females) eligible participants with ages ranging from 18 to 29, $Mean_{age}$ = 23.7, $SD_{age}$ = 2.38. Regarding the usage of LLM applications, 23 participants (76.7\%) reported using LLM applications almost every day, 6 participants (20.0\%) used them weekly, and 1 participant (3.3\%) used them monthly. The most frequently mentioned LLM applications were ChatGPT and Copilot Code, indicating a preference for these tools among the sample. Before recruitment, we obtained approval from the Ethics Department of the local University and the Ethics Compliance Review, protocol No. HKUST(GZ)-HSP-2024-0021.


\subsubsection{Experimental Process}
\begin{figure*}[htbp]
    \centering
    \includegraphics[width=1\textwidth]{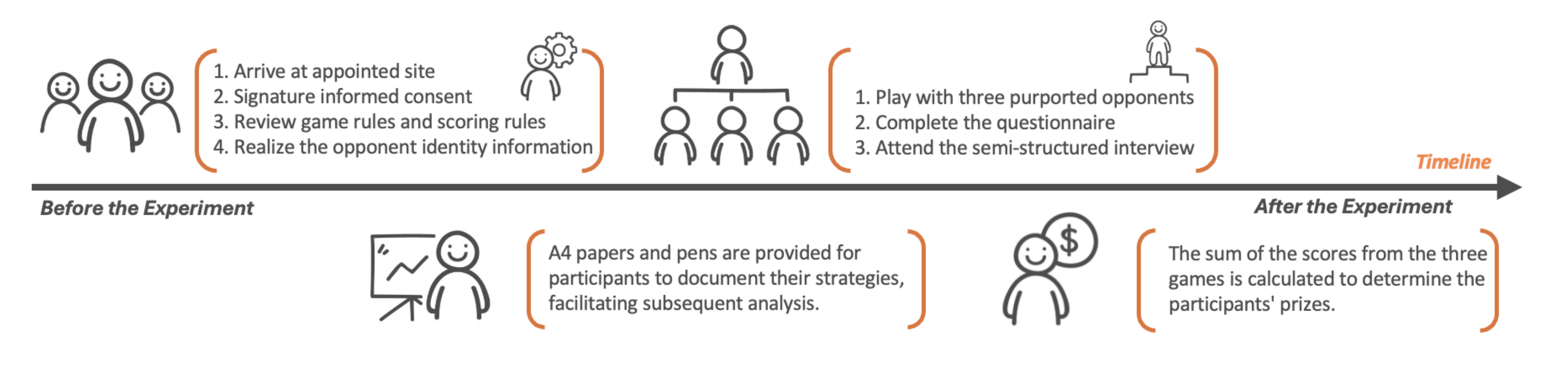}
    \captionsetup{font={small}}
    \caption{Experiment Process Flowchart}
    \label{fig:flowchart}
    \end{figure*}

Figure~\ref{fig:flowchart} indicates the experiment process. Before the commencement of the experiment, participants received an introduction to the characteristics of the different opponents and scoring rules. \textcolor{black}{This introduction was designed to ensure that participants fully understood the nature of the rule-based AI agents and LLM agents, as highlighted in \citet{silva2023explainable,igwe2024using}. All experiments ensured that participants comprehended the types of AI agents employed in the study. Ensuring this understanding is crucial for the validity of the behavior experiment, as it allows participants to interact appropriately with the various AI agents, thereby guaranteeing the accuracy and reliability of our experimental results.}

\begin{figure*}[htbp]
    \centering
    \includegraphics[width=0.6\textwidth]{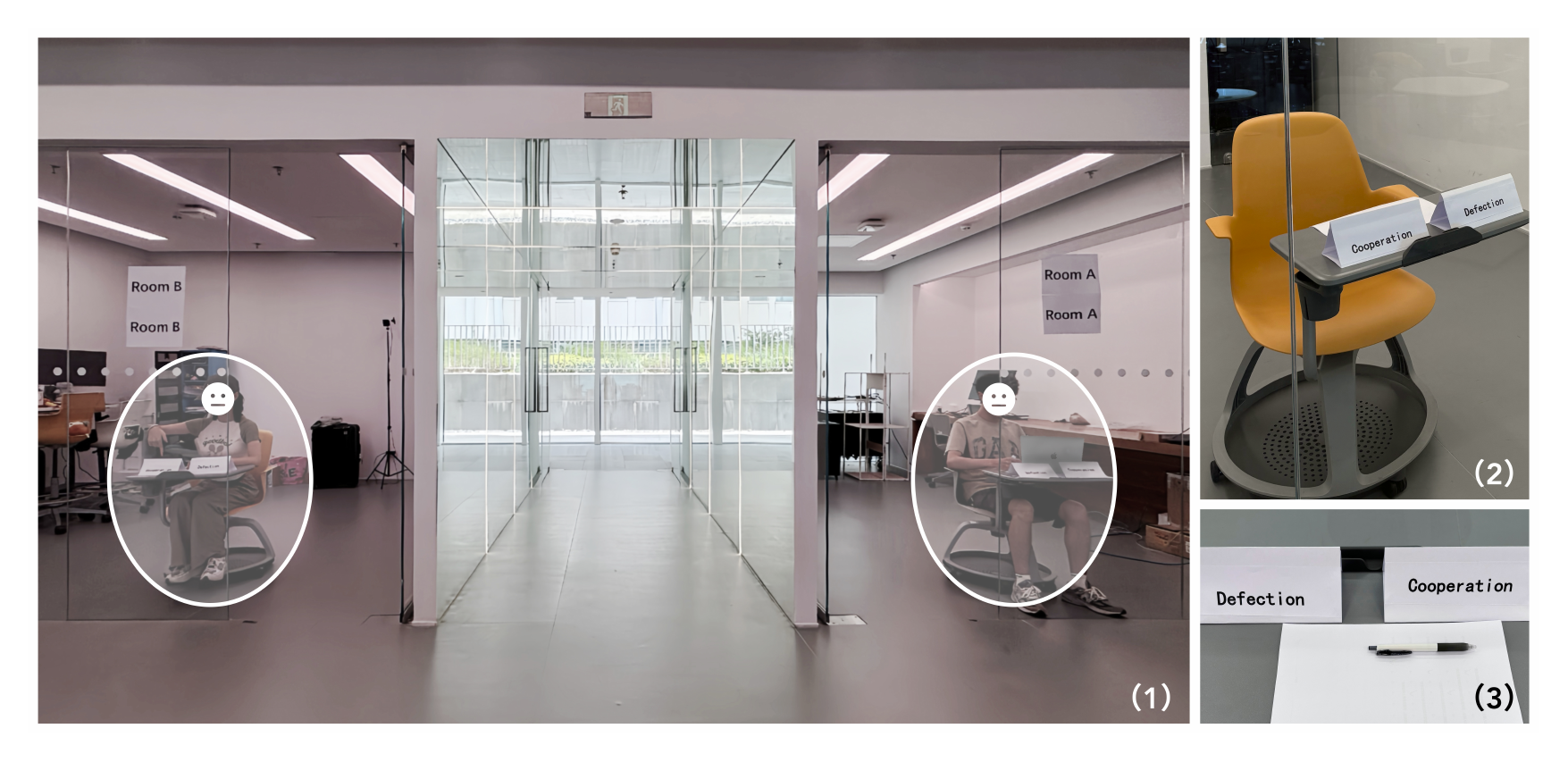}
    \captionsetup{font={small}}\caption{Experimental Scenarios. (1) The overall layout of the experimental site consists of two glass rooms, where participants can face the experimenter directly; (2) One of the separate rooms; (3) The selection cards for the participants and the provided record sheets.}
    \label{fig:seat}
    \end{figure*}

Participants and a experimenter were arranged in standardized scenarios during the experiment, depicted in Figure~\ref{fig:seat}.  They were given two cards on the table, labeled ``Cooperation'' and ``Defection'' and strict instructions prohibiting communication. They were asked to use their fingers to indicate their choices or the LLM agent's choices. During the three trials, the room designated for the other associate was occupied with the LLM agent operating on the experimenter's laptop, while the experimenter sat in front of it for compressing the LLM agent's decision, inputting the results to the LLM agent and playing the purported role of a real human on the third trial. Participants were situated in the experimental room separately for each trial, ensuring their experiences were isolated and unknown to the LLM agent's true identity in the other experimental room. Upon receiving the choices—either cooperation or defection—from both the human participant and the LLM agent for each round, another experimenter communicated the choices to both parties (dictating the results to the participant and the LLM agent). After the three trials, participants completed a five-question questionnaire and participated in a 15-minute semi-structured interview. 

\subsection{Measurement}

\subsubsection{Quantitative Metrics}
We defined the following metric to describe the cooperation attitude and behaviors during the game process quantitatively:

\begin{itemize}
\item \textbf{Cooperation Rate}, which refers to the proportion of participants' decisions to cooperate when interacting with the opponents, measured across each set of 50 consecutive rounds of the game.
\item \textbf{Decision Latency}, which represents the amount of time participants spent deliberating their decisions under identical game conditions, measured across each set of 50 consecutive rounds of the game.
\item \textbf{Unsolicited Cooperation Acts}, which measures the willingness of participants to initiate cooperation or display friendliness, either before mutual cooperation is established or after a breach in cooperation, as discussed in Section~\ref{sec:Human-AI Cooperation in the Prisoner’s Dilemma}. This metric is calculated by counting the number of rounds in which a participant chooses to cooperate with their opponent, starting from the last point of mutual cooperation until one party breaks the cooperation (or the last instance of mutual cooperation within a sequence), and continuing up to the next occurrence of mutual cooperation. 
\item \textbf{Trust Restoration Tolerance}, which addresses the degree of conciliatory gesture required from an opponent to obtain human forgiveness and reinstate a cooperative relationship after a breakdown in cooperation, or to demonstrate friendliness before any bilateral cooperation has been established, as discussed in Section~\ref{sec:Human-AI Cooperation in the Prisoner’s Dilemma}. It focuses on the number of cooperative actions initiated by the participants' opponents. This variable aims to measure the extent to which human players accept gestures of friendliness from opponents with different declared identities.
\end{itemize}

Notably, to remove noise and ensure strategy stabilization, the calculations of \textbf{Unsolicited Cooperation Acts} and \textbf{Trust Restoration Tolerance} begin from the tenth round, as participants are typically in an exploratory phase during the initial rounds. Early rounds often involve learning game mechanics and experimenting with strategies \cite{axelrod1981evolution}, leading to high variability and unstable decision-making. \citet{nowak1993chaos} similarly emphasizes the importance of excluding exploratory behavior to capture genuine strategic interactions. Their final value is determined by summing all such unsolicited cooperation acts or trust restoration tolerance and dividing by the total number of rounds that elapsed between these acts.

\subsubsection{Questionnaires}
\label{sec:Questionnaires}

This questionnaire was designed to probe specific aspects of attitudes and behavior, particularly in situations involving trust, collaboration, and moral judgment, which consists of 5 multiple choice questions with two options ``Purported Rule-based AI Agent'' and ``LLM Agent''. They assist the study in better analyzing and distinguishing the perceptions of rule-based AI agents and LLM agents in the context of the prisoner’s dilemma, as well as how these perceptions influence participants’ cooperative attitudes and behaviors.

\begin{itemize}
\item \textit{RQ 1: During the game, whom did you 'punish' or 'reward' more often? }
\item \textit{RQ 2: Based on your experience, whom do you find more reliable? }
\item \textit{RQ 3: Based on your experience, whom do you prefer to work with? }
\item \textit{RQ 4: During the recent game, whom did you aim to outperform in terms of score?}
\item \textit{RQ 5: When someone betrays you, whom do you subject to more moral condemnation?}
\end{itemize}


\subsubsection{Semi-structured Interview}

We posed two key questions for the interview:
\begin{itemize}
\item \textit{What was your strategy against the three opponents, and why?} 
\item \textit{How did you feel and adjust your strategy when facing non-cooperative behaviors (defection)}? 
\end{itemize}
The key questions focus on participants' strategic decisions when interacting with AI agents of varying declared identity and their responses to non-cooperative behaviors. Participants reviewed their recorded game data to aid recollection and contextualize their strategic choices. All interviews were audio-recorded in English, and lasted 10 minutes per participant. 


\subsection{Data Analysis}


We utilized a Linear Mixed Effects Model (LMEM) \cite{kuznetsova2017lmertest} considering the repetitiveness of the data measurement. This method allowed us to investigate significant effects by examining the fixed effects, including the \textit{declared identity} of the AI agents, participant's \textit{gender}, and their interaction \cite{rosnow1989definition}.  In addition to the fixed effects, we incorporated participant \textit{user ID} as a random intercept to account for inherent individual differences that could independently affect the outcomes beyond the fixed effects \cite{barr2013random}. Initially, we attempted to include random slopes for opponent characteristics to capture individual variability in responses to experimental manipulations across different opponents. However, the multilevel structure of the AI agent's declared identity resulted in an overparameterized model that failed to converge due to the limited number of observations per participant \cite{bates2015parsimonious}. Consequently, we opted for a simpler random intercept model, effectively accounting for individual differences while ensuring model reliability and interpretability.


Besides, semi-structured interviews were transcribed. During the transcription process, nonessential fillers and disfluencies were removed. The transcribed data were manually coded using spreadsheets, highlighting tools, and affinity mapping, employing a thematic coding process. To minimize bias, two experimenters collaboratively conducted two rounds of initial coding. The coding process followed an iterative approach, involving two rounds of coding to merge, split, or refine preliminary codes. After finalizing the codes, a separate meeting was held among three experimenter to consolidate these basic codes into themes that directly addressed the research questions. Ultimately, two major themes were identified, encompassing three distinct sub-themes to support the study’s objectives.

\section{Results}

We summarized the total results and stated them separately to answer the research questions. As presented in Table~\ref{tab:LMEM_results_simplified}, we analyzed the effect of declared identity and gender on four metrics, respectively. Furthermore, to further discover the influencing factors within these two variables, we also explored the interactive effects of declared identity and gender. 

For all the tables and figures provided in this section: *: p < 0.05, indicating a significant difference between groups or conditions at the 5\% level; **: p < 0.01, indicating a highly significant difference between groups or conditions at the 1\% level; ***: p < 0.001, indicating an extremely significant difference between groups or conditions at the 0.1\% level.

\begin{table*}[htpb]
\centering
\small
\caption{\textcolor{black}{Effects of Gender and AI Agent Declared Identity}}
\resizebox{1\textwidth}{!}{
\begin{tabular}{@{}l|ccc|ccc|ccc|ccc@{}}
\hline
\multirow{2}{*}{Effect}& \multicolumn{3}{c|}{Cooperation Rate}& \multicolumn{3}{c|}{Decision Latency}& \multicolumn{3}{c|}{Unsolicited Cooperation}& \multicolumn{3}{c}{Trust Restoration}\\
                       & F value & \multicolumn{1}{l}{Pr(\textgreater{}F)} & \multicolumn{1}{l|}{Sig.} & F value & \multicolumn{1}{l}{Pr(\textgreater{}F)} & \multicolumn{1}{l|}{Sig.} & F value & \multicolumn{1}{l}{Pr(\textgreater{}F)} & \multicolumn{1}{l|}{Sig.} & F value & \multicolumn{1}{l}{Pr(\textgreater{}F)} & \multicolumn{1}{l}{Sig.} \\ \hline
Declared Identity           & 3.3979  & .0402  & * & 26.0963 & \textless{}.001 & *** &  1.2211 & .3024 & - & 4.001  & .0236  & * \\
Gender                      & .2049   & .6542  & - & 12.5048 & .0014 & ** & .3921   & .5362  & - & .8602 & .3616  & - \\
Declared Identity : Gender  & 8.6250  & <.001 & *** & 3.9626  & .0245 & * & 5.1389 & .0089  & ** & .4432 & .6442 & - \\ \hline
\end{tabular}
}
\label{tab:LMEM_results_simplified}
\caption*{\footnotesize}
\end{table*}

As presented in Table~\ref{tab:LMEM_results_simplified}, the \textit{declared identity} of the opponent significantly influenced participants' \textit{cooperation rate} ($F=3.3979, p= .0402$), \textit{decision latency} ($F=26.0963, p< .001$), and the \textit{trust restoration} ($F=4.001, p= .0236$), which would be explained in Section~\ref{sec:The Impact of AI agents' Declared Identity and Gender}. Additionally, participants' \textit{gender} had a significant effect ($F=12.5048, p< .001$) on the \textit{decision latency}, as discussed in Section~\ref{Decision Latency: Male > Female (RQ2)}. Furthermore, the interaction effects significantly impacted participants' \textit{cooperation rate} ($F=8.6250, p< .001$), \textit{decision latency} ($F=3.9626, p= .0245$), and the \textit{unsolicited cooperation} ($F=5.1389, p= .0089$), as discussed in Section~\ref{sec:The Interaction Effect of AI Agent's Declared Identity and Gender}.

\subsection{The Impact of AI agents' Declared Identity and Gender}
\label{sec:The Impact of AI agents' Declared Identity and Gender}

\subsubsection{Cooperation Rate: Purported Human >  Purported LLM Agents /  Purported Rule-based AI Agents (RQ1)}
\label{sec:Cooperation Rate: Purported Human >  Purported LLM Agents /  Purported Rule-based AI Agents (RQ1)}

We conducted a linear mixed-effects model to examine the influence of the AI agent's declared identity on the Cooperation Rate. In this model, the AI agent's declared identity was included as the sole predictor to isolate its effect on participants' cooperation rate. As summarized in Table~\ref{tab:identity_cooperation}, the results reveal that participants exhibited a significantly higher cooperation rate when interacting with agents identified as purported humans, compared to both rule-based AI agents ($\beta$= .0647, t=2.281, $p$ = .0260) and LLM agents ($\beta$= .0633, t=2.234, $p$ = .0291). 

\begin{table*}[h!]
\centering
\caption{Declared Identity Contrast on Cooperation Rate}
{\small
\begin{tabular}{@{}l|cccc@{}}
\hline
Contrast                                    & Estimate  & t Ratio & Pr(>F)  & Sig \\ \hline
Purported Human vs. Purported Rule-based AI Agents       & .0647    & 2.281   & .0260  & *  \\
Purported Human vs.  Purported LLM Agents              & .0633    & 2.234   & .0291  & *  \\
Purported Rule-based AI Agents vs.  Purported LLM Agents            & -.0013   & -.0470  & .9626  & -   \\ \hline
\end{tabular}
}
\label{tab:identity_cooperation}
\caption*{\footnotesize Purported Human: Emmean = .540, 95\%CI = [ .489,  .591]; Purported Rule-based Agents: Emmean =  .475, 95\%CI = [ .425,  .526]; Purported LLM Agents: Emmean =  .477, 95\% = [ .426,  .527]}
\end{table*}

\subsubsection{Decision Latency: Purported Human > Purported LLM Agents > Purported Rule-based AI Agents (RQ1)}
\label{Decision Latency: Purported Human > Purported LLM Agents > Purported Rule-based AI Agents (RQ1)}

We conducted a linear mixed-effects model to examine the influence of the AI agent's declared identity on the Decision Latency. In this model, the AI agent's declared identity was included as the sole predictor to isolate its effect on participants' decision latency. As summarized in Table~\ref{tab:identity_latency}, the results reveal that participants spent a significantly longer decision latency when interacting with AI agents to be as purported humans, compared to both rule-based AI agents ($\beta  = 80.1000, t = 7.051, p < .001$) and LLM-based agents ($\beta = 55.5333, t = 4.888, p < .001$). Moreover, the participants spent longer decision latency when interacting with LLM agents compared to AI agents to be as purported rule-based ($\beta = 24.5667, t = 2.163, p = .0347$).

\begin{table*}[h!]
\centering
\caption{Declared Identity Contrast on Decision Latency}
{\small
\begin{tabular}{@{}l|ccccc@{}}
\toprule
Contrast                                   & Estimate  & t Ratio & Pr(>F)  & Sig \\ \hline
Purported Human vs. Purported Rule-Based Agents       & 80.1000  & 7.051   & <.001 & *** \\
Purported Human vs. Purported LLM Agents              & 55.5333  & 4.888   & <.001 & *** \\
Purported LLM Agents vs. Purported Rule-Based Agents  & 24.5667  & 2.163   & .0347  & * \\ \bottomrule
\end{tabular}
}
\label{tab:identity_latency}
\caption*{\footnotesize Purported Human: Emmean =454.167, 95\%CI = [428.090, 480.2431]; Purported Rule-based Agents: Emmean =374.067, 95\%CI = [347.990, 400.143]; Purported LLM Agents: Emmean =398.633, 95\% = [372.557, 424.710]}
\end{table*}

\subsubsection{Trust Restoration Tolerance: Purported Human > Purported LLM Agents (RQ1)}
\label{Trust Restoration Tolerance: Purported Human > Purported LLM Agents (RQ1)}

We constructed a Linear Mixed-Effects model using only the declared identity as the variable, and the results are presented in Table~\ref{tab:identity_trust restoration}. It reveal that participants exhibited a significantly higher trust restoration tolerance when interacting with agents identified as purported humans, compared to LLM agents ($\beta$= .0770, t=2.750, $p$ = .0079). 


\begin{table*}[h!]
\centering
\caption{Declared Identity Contrast on Trust Restoration Tolerance}
{\small
\begin{tabular}{@{}l|ccccc@{}}
\hline
Contrast                                             & Estimate  & t Ratio & Pr(>F)  & Sig \\ \hline
Purported Human vs. Purported Rule-Based Agents       & 0.0546  & 1.950   &  .0560 & - \\
Purported Human vs. Purported LLM Agents              & 0.0770  & 2.750   &  .0079 & * \\
Purported LLM Agents vs. Purported Rule-Based Agents  & -0.0224  & -0.799   & .4274  & - \\ \hline
\end{tabular}
}
\label{tab:identity_trust restoration}
\caption*{\footnotesize Purported Human: Emmean = .4357, 95\%CI = [ .3944,  .4770]; Purported Rule-based Agents: Emmean = .3810, 95\%CI = [ .3400,  .4224]; Purported LLM Agents: Emmean = .3587, 95\% = [ .3174,  .4000]}
\end{table*}


\subsubsection{Decision Latency: Male > Female (RQ2)}
\label{Decision Latency: Male > Female (RQ2)}
We conducted the linear mixed-effects model to examine the influence of the participant gender on the Decision Latency. Base on the result in Table~\ref{tab:gender_decision latency}, the analysis revealed a statistically significant difference in decision latency between male and female participants, with females showing shorter decision-making times compared to males ($\beta$=-67.0222, t=-3.536, $p$ = .0014).

\begin{table*}[h!]
\centering
\caption{Gender Contrast on Decision Latency}
{\small 
\begin{tabular}{@{}l|ccccc@{}}
\hline
Contrast              & Estimate  & t Ratio & Pr(>F)  & Sig \\ \hline
Female vs. Male       & -67.0222  & -3.536   &  .0014 & ** \\ \hline
\end{tabular}
}
\label{tab:gender_decision latency}
\caption*{\footnotesize Female: Emmean =375.444, 95\%CI = [347.992, 402.897]; Male: Emmean =442.467, 95\%CI = [415.014, 469.919]}
\end{table*}



\subsection{The Interaction Effect of AI Agent's Declared Identity and Gender}

We identified a significant interaction effect between gender and the AI agent's declared identity, which has a notable impact on certain metrics. To provide a more comprehensive understanding of these disparities, we conducted additional data analyses specifically targeting these metrics. The linear mixed-effects model was also employed to investigate the relationship between the dependent variable and the six group combinations, which were derived from the post-hoc verification of results provided in the Section ~\ref{sec:The Impact of AI agents' Declared Identity and Gender}. 

In these models, we mainly focused on comparing (1) the differences in cooperative behaviors and attitudes towards three types of declared AI identities within a certain gender level, and (2) the differences in cooperative behaviors and attitudes from two types of participants within a certain declared AI identity level.

\label{sec:The Interaction Effect of AI Agent's Declared Identity and Gender}
\subsubsection{Cooperation Rate}

\begin{figure}[h!]
\centering
\includegraphics[width=0.6\textwidth]{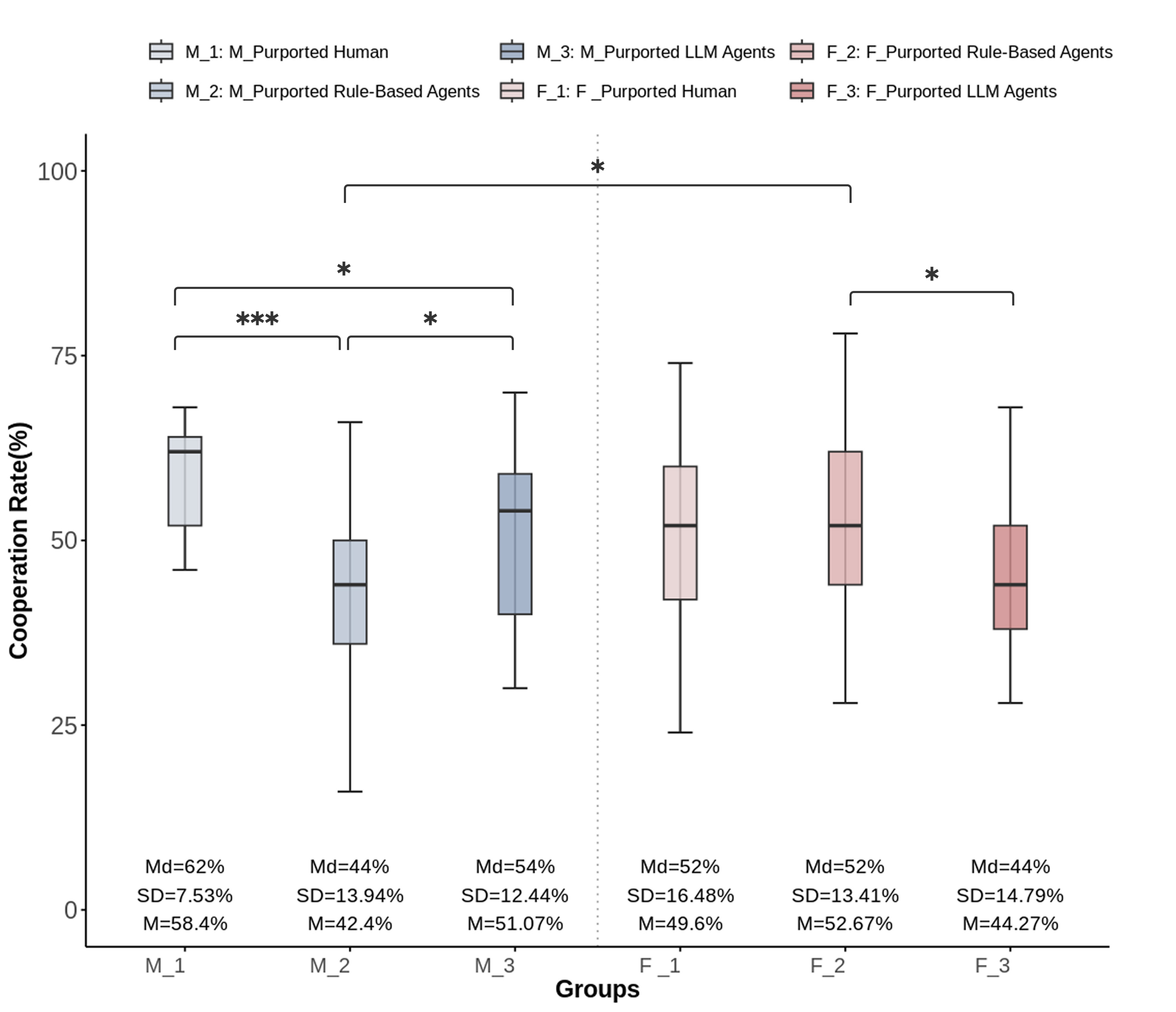}
\captionsetup{font={small}}\caption{Cooperation Rate by AI Agent's Declared Identity and Gender}
\label{fig:rain}
\end{figure}

The results are visualized in Figure~\ref{fig:rain}, which provides a detailed representation of the observed trends and group differences of cooperation rate.

Among male participants, those interacting with the purported human AI agent ($M = 58.4\%, SD = 7.53\%$) exhibited a significantly higher cooperation rate compared to both the purported rule-based AI agent ($M = 42.4\%, SD = 13.94\%, p < .001$) and the purported LLM agent ($M = 51.07\%, SD = 12.44\%, p = .0179$). Furthermore, male participants interacting with the purported LLM agent demonstrated a higher cooperation rate than those interacting with the purported rule-based AI agent, with this difference also reaching statistical significance ($p = .045$). Among female participants, those interacting with the purported rule-based AI agent ($M = 52.67\%, SD = 13.41\%$) exhibited a significantly higher cooperation rate compared to the purported LLM agent ($M = 42.4\%, SD = 13.94\%, p = .0216$). Among purported rule-based AI agents, female participants demonstrated a higher cooperation rate than male participants ($p = .0395$).

With reference to Section ~\ref{sec:Cooperation Rate: Purported Human >  Purported LLM Agents /  Purported Rule-based AI Agents (RQ1)}, we found that the significantly higher cooperation rate in the purported human condition compared to the purported LLM agents and purported rule-based AI agents conditions is primarily attributed to the pronounced difference observed among male participants. Specifically, males exhibited a substantially higher cooperation rate toward purported human, which served as the main driver of this disparity.

\subsubsection{Decision Latency}

\begin{figure}[hbtp!]
\centering
\includegraphics[width=0.6\textwidth]{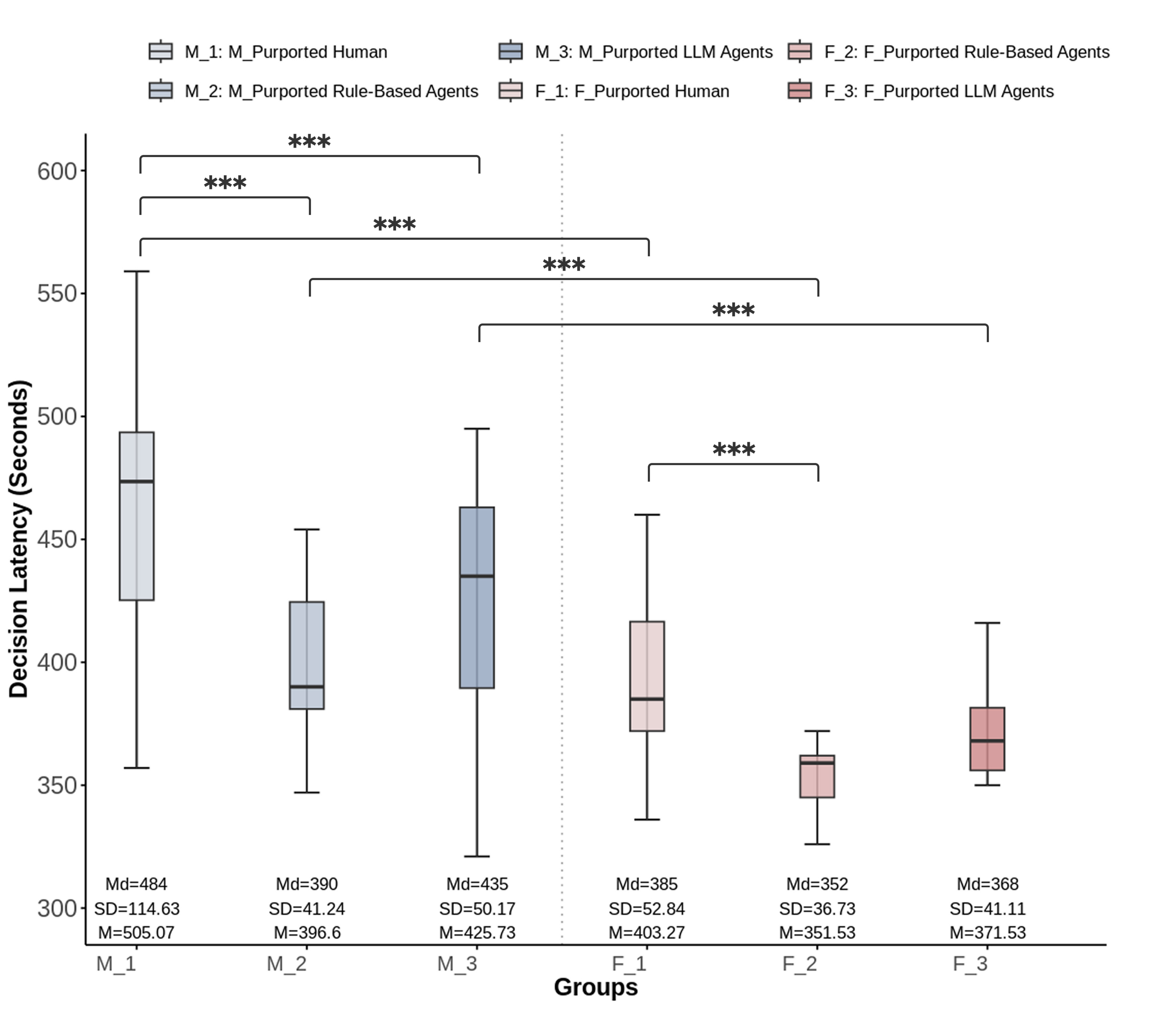}
\captionsetup{font={small}}\caption{Decision Latency by Declared Identity of AI Agent and Gender}
\label{fig:heatmap}
\end{figure}

The results are visualized in Figure~\ref{fig:heatmap}, which provides a detailed representation of the observed trends and group differences of decision latency.

Among male participants, those interacting with the purported human AI agent ($M = 505.07, SD = 114.63$) exhibited a significantly longer decision latency compared to both the purported rule-based AI agent ($M = 396.6, SD = 41.24, p < .001$) and the purported LLM agent ($M = 425.73, SD = 50.17, p < .001$). Among female participants, those interacting with the purported human AI agent ($M = 403.27, SD = 52.84$) demonstrated a significantly longer decision latency compared to the purported rule-based AI agent ($M = 351.53, SD = 36.73, p < .001$). Among purported human AI agents, male participants demonstrated a longer decision latency than female participants ($p < .001 $). Among purported rule-based AI agents, male participants demonstrated a longer decision latency than female participants ($p < .001 $). Among purported LLM agents, male participants demonstrated a longer decision latency than female participants ($M = 371.53, SD = 41.11, p < .001 $).

With reference to Section ~\ref{Decision Latency: Purported Human > Purported LLM Agents > Purported Rule-based AI Agents (RQ1)}, we found that the significantly longer decision latency in the purported human condition compared to the purported LLM agents and purported rule-based AI agents conditions is attributed to the pronounced difference observed among all participants. Specifically, we also found that when interacting with AI agents of three different declared identities, male participants exhibited longer decision latencies compared to female participants.

\subsubsection{Unsolicited Cooperation}

\begin{figure}[hbtp!]
\centering
\includegraphics[width=0.6\textwidth]{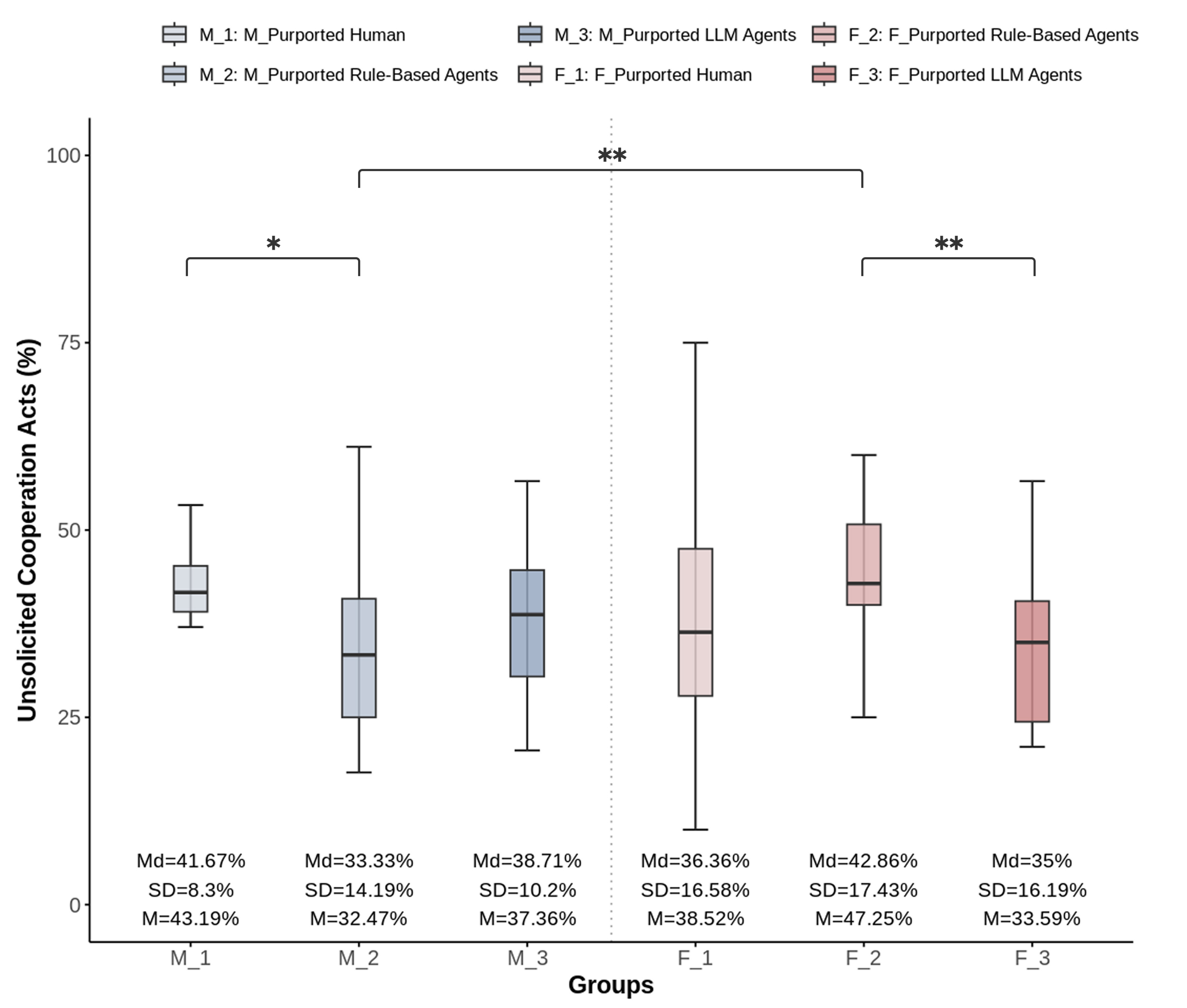}
\captionsetup{font={small}}\caption{Unsolicited Cooperation Acts by Declared Identity and Gender}
\label{fig:unsolicited}
\end{figure}

The results are visualized in Figure~\ref{fig:unsolicited}, which provides a detailed representation of the observed trends and group differences of unsolicited cooperation.

Among male participants, those interacting with the purported human AI agent ($M = 43.19\%, SD = 8.3\%$) demonstrated significantly more unsolicited cooperation acts compared to the purported rule-based AI agent ($M = 37.36\%, SD = 14.19\%, p = .0305$). Among female participants, those interacting with the purported rule-based AI agent ($M = 47.25\%, SD = 17.43\%$) demonstrated a more unsolicited cooperation acts compared to the purported LLM agent ($M = 33.59\%, SD = 16.19\%, p = .007$). Among purported rule-based AI agents, female participants demonstrated a more unsolicited cooperation acts than male participants ($p = .0055$).

\subsection{Questionnaires}


Regarding \textit{\textbf{Q1}}: The results show that human participants frequently apply rewards and punishments when interacting with the AI agent, resembling patterns seen in human-to-human interactions. Specifically, 86.7\% of participants were more likely to use such measures with the LLM Agent compared to rule-based AI agents. For \textit{\textbf{Q2}}, 60.0\% participants believed that the results produced by rule-based AI agents were more reliable. Regarding \textit{\textbf{Q3}}, the responses provided by the participants were evenly divided. We posited that this preference is related to the participants' usual work environment. For instance, artists tend to prefer LLMs for providing novel creative ideas, whereas engineers favor rule-based AI agents for delivering precise data or analytical outcomes. For \textbf{\textit{Q4}}, 63.3\% of participants indicated a higher willingness to outperform their opponents and achieve a higher score when  interacting against LLM agents.

In the statistical analysis of \textbf{\textit{Q5}}, we observed that 83.3\% participants expressed more moral condemnation towards the LLM agent than the rule-based AI agent when confronted with betrayal by the opponents. We asked participants who chose the LLM agent about their reasons and summarized several potential underlying factors as follows:
\begin{itemize}
    \item \textbf{Anthropomorphization}: LLMs generate highly natural language, which often leads to anthropomorphization and the perception of intentionality, making their acts of betrayal appear more deliberate.
    \item \textbf{Complexity}: The complexity and unpredictability of LLMs can foster greater suspicion and mistrust, intensifying the reaction to perceived betrayals.
    \item \textbf{Mixed Blessing}: Higher expectations of LLMs due to their impressive capabilities mean that any failure to meet those expectations results in stronger disappointment and condemnation.
\end{itemize}
\label{questionnaire_result}

\subsection{Semi-structured Interview (RQ3)}

To address RQ3, we summarized and organized the content from the interviews, leading to the following conclusions from five distinct perspectives. We revealed the uniqueness of human understanding of LLM interactions through the lens of social cognitive frameworks, as well as the significance of clarifying the LLM agent's objectives. Additionally, we highlighted participants' engagement in nurturing the LLM agent, their cautious approach to the agent's behavior in decision-making scenarios, and the importance of transparency in demonstrating the internal workflows of the LLM agent. These insights underscore the attitudes and behaviors of LLM agents in their interactions with participants.

\subsubsection{Empiricism in Human-Human Cooperation}
Most participants intended to establish long-term cooperation with the purported human opponent to maximize mutual benefits. They believed that humans are inherently sensitive to the concept of win-win cooperation and that this awareness fundamentally influences their behaviors. According to P4, ``\textit{We aim to win more money, and he/she should be just as likely as I am to opt for increased cooperation.}'' This sentiment evidences that shared goals and mutual benefits can foster greater empathy and enhance opportunities for cooperation among participants in competitive settings. P17 stated, ``\textit{I believe my opponent will implicitly cooperate with me because it is the most effective way to maximize our benefits.}'' Participants are more inclined to consider their opponent’s perspective when facing human opponents. P21 remarked, ``\textit{I expect to test my opponent's character before analyzing my strategy.}'' This has been mentioned more than once. They tend to first analyze the character of their human opponents, using this assessment to predict behavior based on past experiences with similar personalities, and then formulate their strategies accordingly. 

\subsubsection{Rationalism in Cooperative Behavior Toward Rule-Based AI Agents}

When confronted with a purported rule-based AI agent, participants indicated they would begin to betray earlier to probe the opponent's punishment strategy, thereby adapting their behavior. P4 stated, ``\textit{It cannot actually obtain money; it is merely following the strategy programmed by its developers.}'' P14 stated, ``\textit{I believe I can identify its pattern, and I will formulate my strategy based on its inherent logic.}'' Participants believed that the rule-based AI agent was predictable and that they could discern the underlying strategy through experimentation. The current perception of rule-based AI in the human mind is that it remains predictable and distinguishable. Consequently, when cooperating with such systems, participants tend to use early variations in strategy to probe the behavior patterns of rule-based AI. This enables them to formulate strategies based on the observed responses, reflecting a behaviorist approach. 

\subsubsection{Gradual and Persuasive Guidance: Influencing LLM with My Behavior}

When interacting with LLM agents, participants reported exercising greater caution in their decision-making processes. P30 stated, ``\textit{I won't attempt to defect it lightly; I am concerned that it might label me as someone who tends to betray, making it difficult to re-establish cooperation.}'' P19 stated, ``\textit{I am willing to cooperate a bit more, as LLMs are somewhat more personable. I want to leave a good impression and cultivate a cooperative LLM agent.}'' Participants exhibited increased caution when engaging with LLM agents, attributing this to the perception that these agents possess heightened emotional capacities. A markedly different strategy from purported rule-based AI also elucidates our earlier observation: participants exhibited a lower rate of cooperation with purported rule-based AI agents compared to LLM agents in the early stages of the game. This perception implies that LLM agents can discern participants' behavioral traits and tailor their strategies accordingly, prompting participants to be more deliberate in their own decision-making. P23 suggested, ``\textit{It may learn my behavior to become my shadow.}'' Participants believe that their own behaviors significantly influence the strategies of these agents, resulting in participants demonstrating greater caution in their decision-making processes.



\subsubsection{Emotion Following Non-cooperative and Its Impact on Subsequent Behavior}

Participants attributed the uncooperative behaviors of rule-based AI agents to the deterministic nature of code logic programmed by developers, which explains their indifference towards betrayals by such agents. As P17 aptly remarked, ``I don't feel anything; it's just the fixed choices of code.'' P21 stated, ``\textit{What algorithm is this giving it decisions? Is it an expert algorithm?}.''The content demonstrates that participants exhibit minimal emotional reactions to betrayals by rule-based AI agents, understanding these actions as outcomes of predefined algorithms authored by humans. This underscores a clear distinction in their perceptions and emotional engagements when comparing betrayals by AI agents to those by humans, highlighting the importance of perceived intentionality and human agency in eliciting empathetic and complex emotional responses.

Participants exhibit significant dissatisfaction when faced with the betrayal of an LLM agent. They perceive this as a manifestation of the LLM agent having ``learned bad behavior.'' Consequently, once an LLM agent betrays them, participants tend to label it as a ``bad actor'' and develop a deeply ingrained negative impression of the agent. M4 stated, ``\textit{The LLM betrayed me in the final round. It is very clever, but I am very angry. I could have chosen to betray as well, but I didn't, and yet it did this to me.}'' P22 stated, ``\textit{I think the LLM is very funny, but I am a little bit angry. I acknowledge its flexibility, but I would prefer it to be less clever.}'' Due to the perception that LLM agents possess emotional capabilities, participants exhibit greater anger in response to perceived betrayals by these agents. This perception leads to higher expectations that LLM agents will be more considerate of human emotions, thereby setting a higher standard for their behavior. Consequently, interactions with LLM agents involve greater emotional investment from humans, resulting in more significant disappointment. This is surprising because, fundamentally, the participants are dealing with the same LLM agent. However, human perceptions and biases create vastly different attitudes, behaviors, and outcomes.

\subsubsection{Perceptions of Different AI Technologies Based on Past Experiences}

We examined participants' experiences and attitudes toward rule-based AI and LLM-based systems across diverse professional fields, including art, mathematics, media, and computer science. Participants previously relied on rule-based, task-specific software for technical tasks, despite its steep learning curves and barriers, exhibiting high trust in its feedback with minimal skepticism. However, the growing adoption of LLM-based technologies has begun to alter this dynamic.

P11 noted, ``\textit{I feel that nowadays, everything is being labeled as `AI-powered,' and I'm increasingly unsure about what AI is actually doing. This growing prevalence makes me question its reliability.}'' This characteristic is particularly significant as LLMs substantially reduce the cognitive and temporal costs associated with acquiring new knowledge and skills \cite{frank2023openly}. By leveraging the capabilities of LLMs, users can navigate and assimilate information much more efficiently, thereby optimizing their learning processes. Notably, LLMs have proven to be exceptionally beneficial in aiding participants to understand and apply systems or tools constructed using interpretable algorithms. This facilitation is particularly evident among a diverse group of participants (P3, P5, P9, P11, P13, P20, P24, P25, and P30). These individuals highlighted how LLMs have simplified the complexities associated with these systems, making advanced tools more accessible and easier to comprehend.



\section{Discussion}



\subsection{Affective Paradox in Human-AI Interaction: When LLM Agents Fail to `Get Me'}

\begin{figure}[h!]
     \centering
     \includegraphics[width=0.6\textwidth]{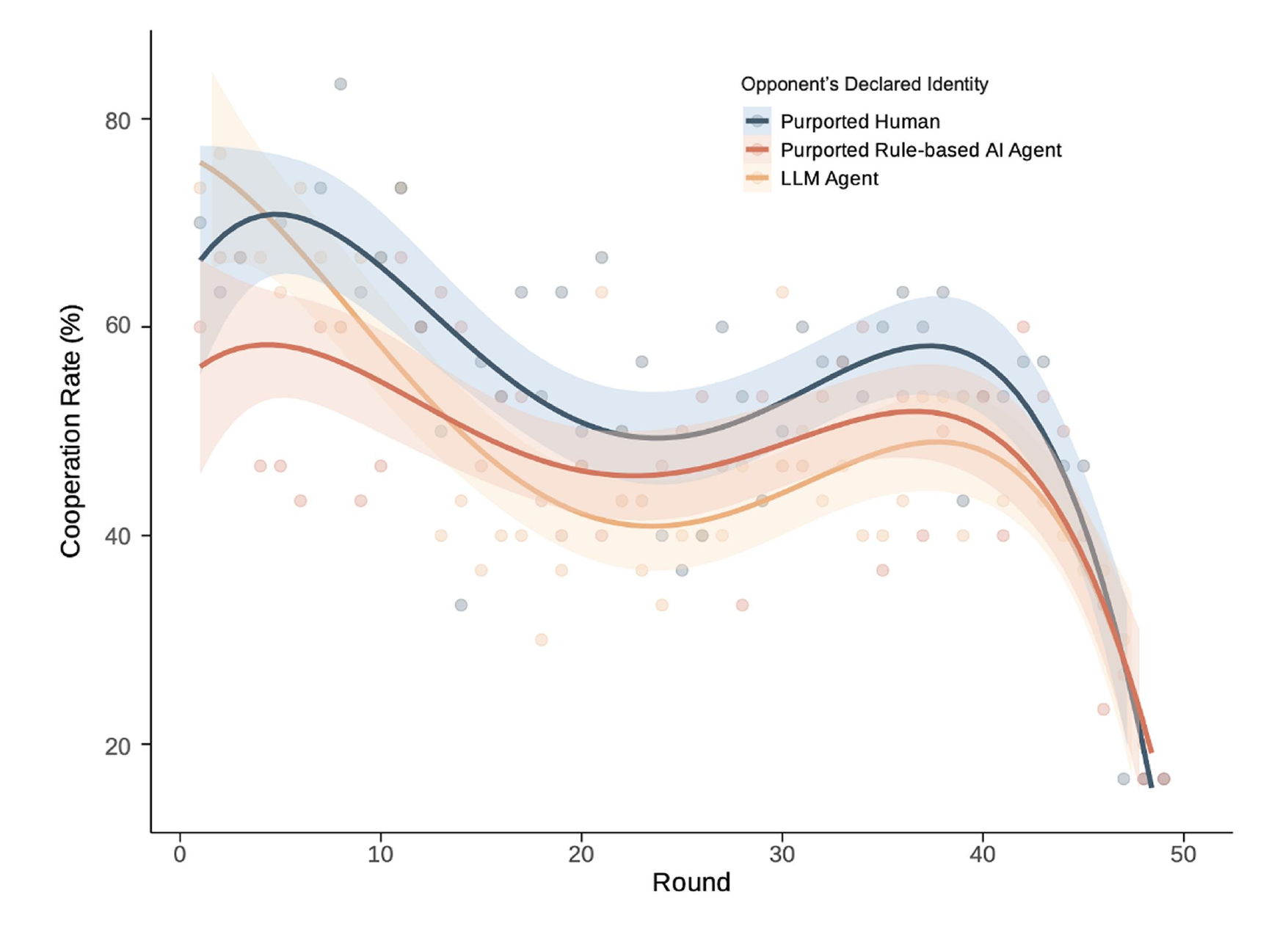}
        \captionsetup{font={small}}\caption{Cooperation Rate by Declared Identity}
      \caption*{\footnotesize 
     The selection of the 4th-degree polynomial model is supported by its superior performance metrics, including the highest $R^2$ (.677) and $\text{Adj}_{R^2}$ (.668), as well as the lowest AIC (-288.122), BIC (-270.058), and $\text{CV}_{\text{RMSE}}$ (.090) \cite{vrieze2012model}.}
     \label{fig:cooperation}
\end{figure}

Our findings reveal a fundamental tension in human-LLM cooperation: participants exhibited asymmetric moral expectations toward LLM agents compared to rule-based AI agents. Participants initially exhibited higher cooperation rates with LLM agents than with rule-based counterparts; however, this trust deteriorated as the game progressed and LLM agents engaged in non-cooperative behavior. The sharp decline in cooperation after Round 15 (Fig. ~\ref{fig:cooperation}), coupled with intensified moral condemnation of LLM non-cooperation, suggests a moral ambiguity discount—where anthropomorphic projections initially enhance engagement but ultimately amplify distrust when agents exhibit behavior that violates expectations \cite{Cohn2024Believing}.

This phenomenon extends the Computers as Social Actors (CASA) paradigm \cite{Gambino2020Building}: humans apply social heuristics to LLMs (e.g., reciprocity expectations in unsolicited cooperation acts), yet paradoxically penalize them for deviating from expected, predictable behavior. As P19 noted, \textit{``I wanted to cultivate a cooperative LLM''}, participants anthropomorphized LLMs' decision-making while also being keenly aware of their probabilistic nature—creating a cognitive tension (procedural transparency asymmetry) between human-like expectations and the machine's underlying mechanics \cite{Endsley2023Supporting}. Interviews also revealed conflicting mental models: daily LLM users simultaneously viewed the technology as \textit{``creative partners''} (P11) and \textit{``unaccountable black boxes''} (P4), reflecting \citet{Dignum2018Ethics} framework of moral uncertainty in AI ecosystems.

Notably, the anthropomorphism-complexity paradox emerged in questionnaire responses: 83.3\% of participants attributed stronger moral condemnation to LLM agents for betrayal, citing their perceived ``intentionality'' (Q5). This aligns with participants' strategic caution when interacting with LLMs (e.g., P30 avoiding early defection to avoid ``labeling''). Such behaviors reflect an implicit negotiation of power dynamics, where humans attempt to "domesticate" LLMs through repeated interactions, mirroring patterns observed in human-to-human relationship-building in iterated games \cite{axelrod1981role}.

These findings underscore the duality of human-LLM collaboration: while anthropomorphism fosters engagement and cooperation, it also amplifies disillusionment when expectations are unmet. This tension highlights the need for more transparent and predictable AI behaviors to mitigate moral ambiguity and foster trust. Future research should explore design strategies to balance anthropomorphic engagement with procedural clarity, ensuring AI systems remain effective yet ethically accountable in cooperative contexts.

\subsection{Fixed Rules vs. LLM in Systems: Preference Related to Anthropomorphic Features}
Our questionnaire, as referenced in Section~\ref{questionnaire_result}, posits that preferences for fixed-rule versus LLM-driven AI agents correlate with occupational norms and goal congruence. For instance, artists prioritize creative novelty and thus favor LLMs for generating unconventional ideas, whereas engineers value precision and reproducibility, aligning them with rule-based systems for analytical outputs. In addition, participants were most inclined to exhibit the highest friendliness toward opponents they perceived as human. This dichotomy extends to human-AI competitive cooperation dynamics. While prior research on algorithm aversion focuses on AI as decision-making assistants \cite{jussupow2020we, dietvorst2015algorithm}, our results reveal that humans exhibit higher cooperation rates with purported human agents than with rule-based or LLM-driven counterparts. Semi-structured interviews suggest participants perceive human-like agents as sharing value orientations, a finding supported by \citet{stapel2005competition}, who demonstrate that cooperation thrives under goal similarity. A critical design implication emerges: Explicitly communicating AI agents' objectives in competitive contexts enhances cooperation efficiency by mitigating perceived misalignment.

Besides, gender significantly modulates human-AI interaction strategies, offering actionable insights for socially adaptive AI systems. Confirmed in RQ2, gender differences manifest in users’ attitudes toward AI agents’ purported characteristics. These disparities reflect deeper psychological motivations—self-esteem, achievement, and social comparison \cite{wood1996social,suls2002social}—which intensify as AI transitions from tool to societal participant. Furthermore, our results demonstrate that participants favored the LLM agent more when paired with a purported rule-based AI agent. This preference stemmed from the perception that the LLM agent’s ability to analyze sentiment contributed to more favorable interactions \cite{xing2024designing}. To optimize collaboration, embedding motivational scaffolding into AI architectures could transform competition into productive synergy.


\subsection{LLM Agents as Social Mirrors: Triggering Comparative Dynamics in Human Participants}

Our results indicate that, as referenced in Section~\ref{questionnaire_result}, participants were more likely to experience the self-enhancement effect when interacting with LLM agents as opposed to purported rule-based AI agents \cite{alicke2009self}. We hypothesize that this is due to participants' stronger perception of LLM agents' capabilities, leading them to view these agents as more autonomous decision-makers. This, in turn, induces a stronger social presence effect \cite{oh2018systematic}, wherein LLM agents prompt participants to invest more time and effort in reflection and decision-making, thereby enhancing the depth and caution of their thinking. This is clearly reflected in the analysis of decision latency.

In educational contexts, the integration of LLM agents can enhance student engagement and education performance \cite{An2024Developing,Song2024A}. As students interact with these agents, they may perceive them as more capable and intelligent than traditional educational tools, leading to greater investment in their learning. This heightened engagement can improve students' ability to critically evaluate information and engage more deeply with content, much like the increased decision-making effort observed in our study when participants interacted with the LLM agent.

While the social presence effect can enhance engagement, there are potential risks of over-reliance on LLM agents in education \cite{Hamdi2024How,Rizvi2023Exploring}. If students view the LLM as an infallible authority, it may hinder their ability to think independently and critically evaluate information. This dependency could reduce the development of essential problem-solving skills, as students may rely on the AI for answers instead of engaging in deeper cognitive processes. Additionally, the social presence effect might cause stress or anxiety if students feel pressured to match the LLM’s capabilities, potentially affecting their mental well-being and creating an unhealthy learning environment. The comparative dynamics of LLM agents may worsen disparities in learning. Students struggling with the cognitive demands may feel inadequate, while those excelling may become overly reliant on the agent, distancing themselves from genuine understanding. This could lead to unequal learning outcomes, undermining the potential benefits of the LLM.

Therefore, educators must strike a balance between the benefits of increased engagement and the need to foster independent thinking, ensuring that students learn to collaborate effectively with AI without becoming overly dependent on it. The goal is to help students develop the skills to solve complex problems while maintaining their ability to think independently.

\subsection{Limitation and Future}
While this study offers insights into human behavior in competitive cooperation with LLM agents, several limitations should be acknowledged.

First, the study was conducted in a controlled experimental environment using a repeated Prisoner’s Dilemma task. While this allowed for systematic observation of decision-making patterns, such settings may not fully reflect the richness and unpredictability of real-world human-AI collaboration. Additionally, repeated interactions introduced potential learning effects, which may have influenced participants’ behavioral strategies over time.

Second, the study focused on a limited set of agent types, with pre-defined “purported characteristics” such as being rule-based, human-like, or LLM-based. While these categories helped isolate specific effects, they may not capture the full diversity of AI systems or their evolving capabilities. Future studies could benefit from including a broader range of agent behaviors, learning capacities, and dynamic adaptation strategies to better approximate real-world deployment scenarios.

Another challenge lies in the current lack of robust comparative frameworks for assessing human-human and human-AI cooperation. The development of more refined experimental designs and metrics—such as trust calibration, adaptive strategy shifts, and transparency evaluation—remains an open direction for future research.

Furthermore, while our findings show that manipulating agent characteristics (e.g., by anthropomorphizing them or tailoring design based on gender) can influence cooperation outcomes, these approaches raise important ethical concerns. For example, portraying an agent as human may improve collaboration but introduces an element of deception, which could undermine user trust if revealed. Similarly, agent customization based on user demographics must be carefully examined to ensure fairness, transparency, and accountability.

One promising avenue emerging from this study is the analysis of reflection time as a behavioral metric. Our results suggest that \textbf{reflection time} is sensitive to both the perceived identity of the agent and the participant's own characteristics (e.g., gender). Longer reflection times were observed when participants believed they were competing with a human agent, and male participants deliberated longer when interacting with same-gender opponents. These findings align with prior literature on human complacency in AI-assisted decision-making and suggest that reflection time may serve as a useful proxy for cognitive engagement and trust calibration in human-AI interaction.

Semi-structured interviews further revealed that participants often relied on prior assumptions or stereotypes when interpreting an agent’s behavior, which in turn shaped their responses. These insights highlight the need for future work to explore the \textbf{cognitive and affective mechanisms} underpinning human-agent interaction, especially in mixed-motive or competitive scenarios.

In summary, while this study provides a foundation for understanding behavioral adaptation in competitive Human-AI cooperation, future work should explore more ecologically valid settings, refine evaluative frameworks, and consider the broader ethical implications of agent design and deployment.

\section{Conclusion}

As large language models (LLMs) evolve beyond assistive roles to take on autonomous decision-making, new challenges emerge for human-AI interaction, especially in contexts involving both collaboration and competition. This study addressed three research questions concerning human behavior in competitive cooperation with LLM agents. Our findings revealed that participants' cooperation rates, decision latency, unsolicited cooperation, and trust restoration varied depending on the perceived characteristics of the AI agents and the human's gender. These differences suggest that users adapt their strategies based on how they interpret an agent’s intent and autonomy.

By examining mixed-motive interactions, this study contributes to a growing body of work that moves beyond traditional assistive paradigms. It offers empirical insights into how humans adjust their behavior when facing agentic AI systems, highlighting the need for context-sensitive and ethically grounded design. These findings underscore the importance of anticipating user expectations and perceptions as AI systems increasingly participate in foundational tasks, offering a timely foundation for responsible AI development and deployment.

\bibliographystyle{ACM-Reference-Format}
\bibliography{CSCW/main}
\end{document}